\documentclass[jcp,twocolumn,superscriptaddress,showpacs,floatfix]{revtex4}
\newcommand{\bpol}{E_b}
\newcommand{\lpol}{l}

\newcommand{\om}{\omega}

\newcommand{\omi}{\om_{i0}}

\newcommand{\eps}{\epsilon}
\newcommand{\epsi}{\eps_{\infty}}
\newcommand{\epss}{\eps_{s}}
\newcommand{\tad}{\tau_{D}}
\newcommand{\tal}{\tau_{L}}
\newcommand{\epom}{\eps (\om)}
\newcommand{\xp}{x^{\prime}}
\newcommand{\wfx}{\psi(x)}

\newcommand{\wfix}{\psi_{i}(x)}
\newcommand{\wfox}{\psi_{0}(x)}

\newcommand{\potx}{\phi(x)}
\newcommand{\potxp}{\phi(\xp)}

\newcommand{\avg}[1]{\langle #1 \rangle}
\newcommand{\norm}[1]{| #1 |^{2}}
\newcommand{\aB}{a_{B}}
\newcommand{\est}{\epsilon^{*}}
\newcommand{\Ry}{\mathrm{Ry}}

\newcommand{\Ei}{E_{i}}
\newcommand{\Eio}{E_{i0}}
\newcommand{\Eoo}{E_{10}}
\newcommand{\Eo}{E_{0}}

\newcommand{\fio}{f_{i0}}
\newcommand{\pio}{\langle \wfix | \partial/\partial x | \wfox \rangle}

\newcommand{\fom}{f(\om)}
\newcommand{\tpol}{\Pi(k)}

\newcommand{\Etot}{E^{\mathrm{tot}}}
\newcommand{\Epol}{E_{\mathrm{P}}}
\newcommand{\potxt}{\phi(x,t)}
\newcommand{\denxt}{\rho(x,t)}

\newcommand{\den}{\bar{\rho}}
\newcommand{\pot}{\bar{\phi}}
\newcommand{\denn}{\rho_{n}}
\newcommand{\denm}{\rho_{m}}
\newcommand{\potn}{\phi_{n}}
\newcommand{\lan}{\lambda_{n}}
\newcommand{\sumn}{\sum_{n \geq 1}}
\newcommand{\ksit}{\xi (t)}
\newcommand{\ant}{a_{n} (t)}
\newcommand{\an}{a_{n}}

\newcommand{\xrelt}{x-\ksit}

\newcommand{\tp}{t^{\prime}}
\newcommand{\taln}{\tau_n}
\newcommand{\kB}{k_{B}}
\newcommand{\fm}{\eta}

\usepackage{graphicx}
\usepackage{amsmath}
\usepackage{latexsym}

\begin{document}

\title{Optical absorption from solvation-induced polarons on nanotubes}

\author{G.~L.~Ussery}
\affiliation{Department of Physics, The University of Texas at
Dallas, P. O. Box 830688, EC36, Richardson, Texas 75083, USA}
\author{Yu.~N.~Gartstein}
\affiliation{Department of Physics, The University of Texas at
Dallas, P. O. Box 830688, EC36, Richardson, Texas 75083, USA}

\begin{abstract}
When an excess charge carrier is added to a one-dimensional (1D) semiconductor immersed in a polar solvent, the carrier can undergo self-localization into a large-radius adiabatic polaron. We explore the local optical absorption from the ground state of 1D polarons using a simplified theoretical model for small-diameter tubular structures. It is found that about 90\% of the absorption strength is contained in the transition to the second lowest-energy localized electronic level  formed in the polarization potential well, with the equilibrium transition energy larger than the binding energy of the polaron. Thermal fluctuations, however, cause a very substantial -- an order of magnitude larger than the thermal energy --  broadening of the transition. The resulting broad absorption feature may serve as a signature for the optical detection of solvated charge carriers.
\end{abstract}


%\date{\today}
\maketitle

\section{Introduction}

One-dimensional (1D) semiconductor (SC) nanostructures (e.~g., nanotubes and nanowires) in contact with polar solvents is an interesting class of systems of particular relevance to applications and processes involving fundamental redox reactions \cite{SCelectrodes,kamat07,pyang05}. When an excess charge carrier (an electron or a hole) is added to such a structure, the carrier can become solvated resulting in a polaron, a self-consistent combination of a localized electronic state and a  dielectric polarization pattern of the surrounding medium \cite{basko,YNGpol, polcylinder,MG_abinitio1,GU_lowfreq}. The long-range Coulomb mechanism of the polaron formation here is analogous to the well-known three-dimensional (3D) polarons in polar SCs \cite{polarons1,polarons2,CTbook} and solvated electrons in polar liquids \cite{CTbook,ferra91,nitzan}.  Distinct from those cases, however, are the confinement of the electron motion in a 1D nanostructure and its structural separation from the 3D polarizable medium.

The physical properties of polarons are quite different from the band electrons frequently discussed in the context of electronic transport in nanotubes \cite{NTelectronics,NTEffectMass} and nanowires \cite{NWelectronics}. We have already emphasized the energetic significance of the solvation by finding \cite{YNGpol,polcylinder} that the binding energy of the resulting 1D polarons could reach a substantial fraction, roughly one third, of the binding energy of Wannier-Mott excitons, the well-known primary photoexcitations in many 1D SCs. This may lead to enhanced charge separation. On the other hand, the mobility of solvated charge carriers is drastically reduced due to the dissipative drag of the medium \cite{basko,CBDrag,GU_lowfreq}. In this paper we address qualitative theoretical expectations for the local optical absorption from solvation-induced 1D polarons. This kind of the optical absorption has been a powerful tool in optical detection of electron-lattice polarons in conjugated polymers as well as of 3D solvated electrons.

While the polaronic effect and the features we discuss have a generic character, specific illustrative calculations in this paper are done using a simplified model representation for small-diameter tubular structures. A widely known example of such structures is semiconducting single-wall carbon nanotubes (SWCNTs). We note that redox chemistry of CNTs has been deemed an ``emerging field of nanoscience'' \cite{chirsel} and solvatochromic effects in CNTs are being intensely researched \cite{CNTsolvchrom}.

\begin{figure}
\includegraphics[scale=1.1]{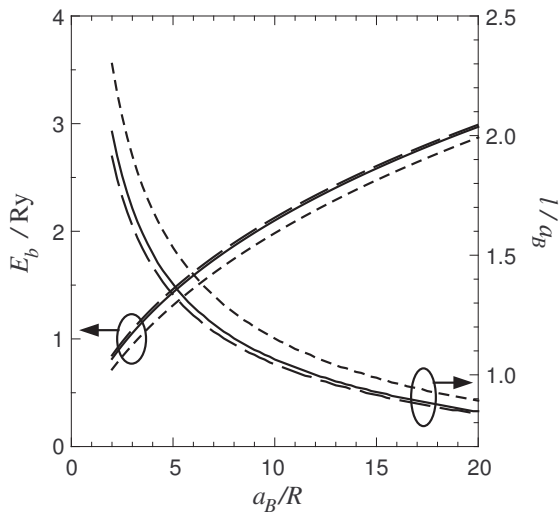}
\caption{\label{pszBE} The binding energy, $\bpol$, and the linear size, $\lpol$, of 1D adiabatic polarons as a function of the confinement parameter $\aB/R$. The size $l$ is defined here as the FWHM of the localized electron density. Different curves correspond to different dielectric conditions, as explained more fully in the text: filled (solid lines), hollow (long-dash), and polarizable (short-dash) nanotube screening models.}
\end{figure}

As discussed in more detail later, Fig.~\ref{pszBE} displays calculated functional dependences for the binding energy, $\bpol$, and the spatial extent, $\lpol$, of 1D adiabatic polarons in terms of the tube radius $R$ and convenient scales of the Bohr radius and Rydberg energy for the corresponding 3D Coulomb problem \footnote{Our definitions here are based on the electron mass rather than on the exciton reduced mass used in Refs.~\cite{YNGpol,polcylinder}.}:
\begin{equation}\label{defaB}
\aB=\est \hbar^2 / m q^2, \ \ \ \ \Ry=q^2/2\est\aB.
\end{equation}
Here $q$ is the carrier charge, $m$ its effective mass, and $\est$ the effective dielectric constant of the uniform 3D medium. The length $\lpol$ has been defined as the full-width-at-half-maximum (FWHM) of the localized 1D electron charge \textit{density}. In what follows we refer to the ratio $\aB/R$ as the confinement parameter. It is well known that the spatial confinement generally results in amplification of the Coulomb binding \cite{haugbook}, and Fig.~\ref{pszBE} illustrates both the magnitude and the growth of $\bpol/\Ry$ for 1D polarons with the increased confinement parameter. For a proper perspective, one should compare \cite{polcylinder} those results with the polaron binding energy in higher dimensions, where the classic Pekar's result \cite{pekar1,appel} for $\bpol$ in 3D  is only $\simeq 0.1 \,\Ry$, while in 2D systems \cite{pol2d} it increases  to $\simeq 0.4 \,\Ry$. Our variational calculations \cite{polcylinder} have shown that the transition to a 1D polaron structure (charge uniformly distributed over the tube circumference) takes place at $\aB/R \gtrsim 1$.

The effective dielectric constant in Eq.~(\ref{defaB}) varies for different solvents and can be determined from the well-known \cite{polarons1,CTbook,nitzan,appel} relationship
$$ %\begin{equation}\label{defest}
1/\est = 1/\epsi - 1/\epss;
$$ %\end{equation}
typical \cite{fawcett,YNGpol} static $\epss$ and high-frequency $\epsi$ constants satisfy $\epss \gg \epsi$ so that $\est \simeq \epsi$. For SWCNTs, effective band masses $m$ of carriers generally depend both on tube radius and chirality; see, e.g., Ref.~\cite{NTEffectMass} for a compilation of some theoretical data. If, for instance, one were to choose a set of representative numerical parameters: $m=0.05 m_e$ ($m_e$ being the free electron mass) and $\est=3$, then Eq.~(\ref{defaB}) results in  $\aB = 32$ {\AA}, $\Ry = 76$ meV, and, with a roughly estimated $\aB/R \sim 4$, $\bpol \sim 0.1$ eV. Along with results demonstrating overall scaling with the confinement parameter,  we will be using specific values of $\aB/R= 4$ and 10 for some plots. One should understand that those are meant to serve mainly illustrative purposes.

\begin{figure}
\includegraphics[scale=1.1]{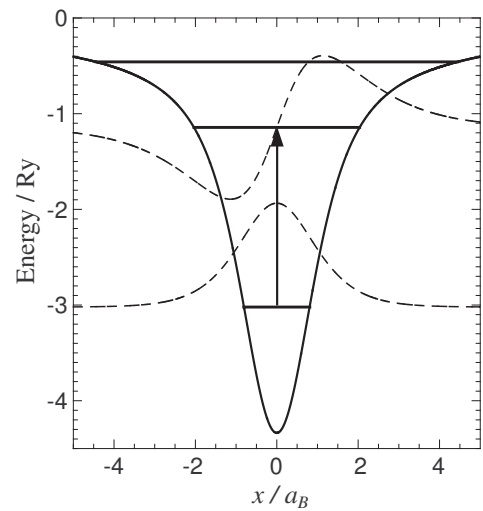}
\caption{\label{pwell} The equilibrium self-consistent 1D potential well ($x$ is a coordinate along the tube axis) along with the three lowest-energy electronic energy levels calculated for $\aB/R=4$. The optical transition with the largest oscillator strength is indicated by the arrow, the dashed lines showing the localized electronic wave functions of the participating states.}
\end{figure}

Figure \ref{pwell} displays an example of the equilibrium 1D polarization potential well calculated for $\aB/R=4$. The corresponding pattern of the medium polarization is self-consistently stabilized by the electric field of a charge carrier occupying the lowest-energy localized electronic level formed in this well (the ground state of the polaron). Along with the lowest, also shown are the two next higher-energy empty levels. The results discussed below indicate that it is the transition between the two lowest-energy levels that has the largest oscillator strength and dominates the optical absorption in the ground state. The long-range behavior of the potential well is Coulombic, and hence it contains many more higher-energy localized states (not shown in Fig.~\ref{pwell}) converging to the onset of the continuum spectrum at zero energy in the figure.

Our consideration in this paper assumes that the polaron binding energy $\bpol$ is sufficiently larger than the thermal energy $\kB T$ so that the thermal occupation of higher-energy electronic states can be neglected. Even so, thermal fluctuations are found to very significantly broaden the optical transition in question, as is also the case for 3D solvated electrons \cite{hydelec}. In the presence of the polaron, the fluctuations involve combined dynamics of the electronic charge density and the solvent dielectric polarization. A major contributor to the broadening is the resulting 1D polaron ``breathing'' mode, qualitatively corresponding to fluctuations of the polaron length. Without pursuing the exact calculation of the absorption line shape, we will provide an approximate estimation of the broadening due to the breathing mode within the classical Franck-Condon framework \cite{CET2004}. The described broad absorption feature could be used for detection of the solvation-induced 1D polarons and an estimate of their binding energy in prospective experimental studies.

\section{The ground-state and optical transitions in equilibrium}

In the single-particle 1D continuum adiabatic framework, the excess charge carrier is described by the wave function $\wfx$; it responds to the electrostatic potential $\potx$ via the corresponding Schr\"{o}dinger equation:
\begin{equation}\label{RSch}
-\frac{\hbar^{2}}{2m}\frac{\partial^{2}\wfix}{\partial x^{2}}+q\potx\wfix=\Ei\wfix.
\end{equation}
In the ground state, the charge carrier occupies the lowest energy ($i=0$) state so that the resulting 1D electronic charge density is
\begin{equation}\label{RRho}
\rho(x)=q\norm{\wfox}.
\end{equation}
At \textit{equilibrium}, the self-consistency requires that this charge density stabilizes the very electrostatic potential used in Eq.~(\ref{RSch}):
\begin{equation}\label{RPot}
\potx=\int d\xp G(x-\xp) \rho(\xp),
\end{equation}
where kernel $G$ is the appropriate electrostatic response function assuming the translational invariance along the $x$-axis. Solving Eqs.~(\ref{RSch})--(\ref{RPot}) together yields the equilibrium 1D polaron distributions that we will be denoting as $\pot (x)$ and $\den (x)$.  One notes that the equations determining the equilibrium correspond to the minimum of the total adiabatic energy of the system:
\begin{equation}\label{ad1}
\Etot_{0}=\Eo + \Epol,
\end{equation}
consisting both of the electronic energy $\Eo$ in Eq.~(\ref{RSch}) and the energy stored in the dielectric polarization of the medium:
$$ %\begin{equation}\label{ad2}
\Epol = \frac{1}{2} \int dx\, d\xp \, \potx \,G^{-1} (x-\xp) \,\potxp,
$$ %\end{equation}
where $G^{-1}$ is the kernel inverse to $G$ (see also Ref.~\cite{YNGpol}). As the continuum onset in Eq.~(\ref{RSch}) is set at zero energy, the ``depth'' of this minimum is the polaron binding energy:
\begin{equation}\label{ad3}
\min\{\Etot_{0}\}=-\bpol.
\end{equation}
The data for $\bpol$ and equilibrium $\den (x)$ and $q\pot (x)$ have been used  in Figs.~\ref{pszBE} and \ref{pwell}. Other adiabatic energy surfaces $\Etot_i=\Ei + \Epol$ are defined similarly to Eq.~(\ref{ad1}).

The electrostatic potential $\potx$ in Eq.~(\ref{RSch}) is due to only the slow (orientational) polarization of the surrounding medium and hence should not include the ``instantaneous'' self-interaction of the charge carrier with itself. A proper separation of the slow polarization response is a standard step in the polaron problem \cite{appel,CTbook,nitzan,CET2004}. The kernel $G$ in Eq.~(\ref{RPot}) can therefore be written as
\begin{equation}\label{resp1}
G(x)=G_s (x) - G_{\infty} (x),
\end{equation}
where indices $s$ and $\infty$ denote that the standard electrostatic potential problem (\ref{RPot}) is solved with dielectric conditions corresponding respectively to full (static, $s$) screening or to screening only by ``fast'' ($\infty$) components of polarization. In the polaron context, ``fast'' means operative on time scales shorter than $\hbar/\bpol$. In the model we consider, the continuum solvent is characterized by the corresponding dielectric constants $\epss$ and $\epsi$. Another ``fast'' contributor to the screening is the polarizability of the 1D SC nanostructure itself related to interband electronic transitions.

In Fourier space, $g(k)=\int dx\, e^{-ikx}\,G(x)$, Eq.~(\ref{resp1}) can be conveniently re-written as
\begin{equation}\label{Gk}
g(k)=g_{0}(k) \left( \frac{1}{\epss(k)}-\frac{1}{\epsi(k)} \right),
\end{equation}
where the $k$-dependence of the corresponding dielectric functions reflects a possible spatial dispersion of the screening due to the geometry of the system.

The bare, unscreened, response for the 1D tubular geometry (the ``elementary'' charge distributions are uniform rings of radius $R$) is given by \cite{ando,YNGpol}
$$ %\begin{equation}\label{Gbare}
g_{0}(k)=2I_{0}(kR)K_{0}(kR),
$$ %\end{equation}
where $I(x)$ and $K(x)$ are the modified Bessel functions appearing in electrostatic problems with cylindrical symmetry \cite{jackson}. The general form of the corresponding $k$-dependent dielectric functions can be represented \cite{YNGpol} as
$$ %\begin{equation}\label{epsk}
\eps (k)=kR\left[\eps_1\,I_{1}(kR)K_{0}(kR) + \eps_2\,I_{0}(kR)K_{1}(kR)\right] + \tpol,
$$ %\end{equation}
where $\eps_1$ is the dielectric constant of the dielectric medium inside the tube, $\eps_2$ the dielectric constant of the unbounded medium outside the tube, and $\tpol$ the contribution arising from the polarizability of the tubular surface itself.

If the latter contribution is neglected and the media inside and outside are the same (we will be calling this model case ``filled''), there would be no spatial dispersion in the screening (uniform medium) and response (\ref{Gk}) takes a simple form of
\begin{equation}\label{Gsimp}
g(k)=-\,g_{0}(k)/\est.
\end{equation}
In order to evaluate the magnitude of the effects arising from different dielectric screening conditions, we also examine the model cases called ``hollow'', where the interior of the tube does not contain any medium ($\eps_1 = 1$), and ``polarizable'', in which the ``hollow'' model is augmented by the tube polarizability. For the model of the latter, we will use the following approximate representation of the polarizability of semiconducting SWCNTs calculated in Ref.~\cite{jiang2007}:
$$ %\begin{equation}\label{polmod}
\tpol=\frac{6.2(kR)^{2}I_{0}(kR)K_{0}(kR)}{1+1.6(kR)^{1.8}}.
$$ %\end{equation}

\begin{figure}
\includegraphics[scale=1.1]{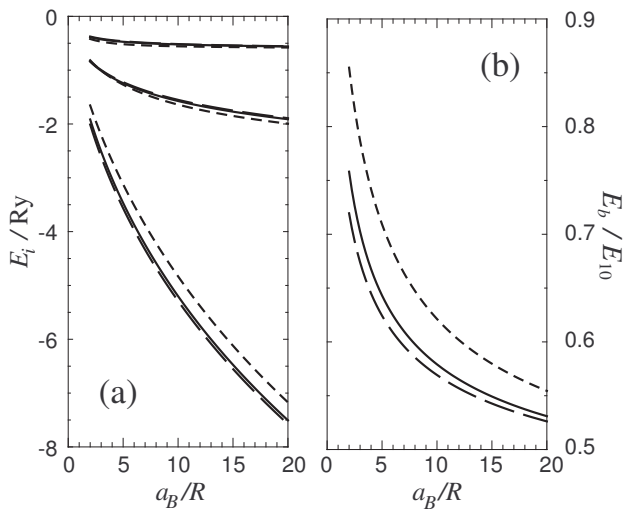}
\caption{\label{eratelev} (a) The energies $\Ei$ of the three lowest electronic energy levels in the ground state equilibrium as a function of the confinement parameter for filled (solid), hollow (long-dash), and polarizable (short-dash) screening models. (b) The ratio of the polaron binding energy, $\bpol$, to the electronic transition energy, $E_{10}$, in equilibrium calculated with the same models.}
\end{figure}

Figure \ref{pszBE} compares the results for the ground state of the polaron  calculated with the three dielectric screening models, in which we used representative values of $\epsi =3$ and $\epss = 40$ for the solvent medium. It is evident from the comparison that in the parameter range we study, different screening models do not lead to drastically different results. The variations observed are quite understandable and are largely caused by the effective changes in the magnitude of the fast screening, which is decreased by going from the filled to hollow model and is increased in the polarizable model. Relatively small changes are also observed in the results for individual electronic energy levels displayed in Fig.~\ref{eratelev}. Panel (a) of that figure just shows the scaling of the three lowest electronic energy levels $\Ei$ ($i=0,1,2$) from Eq.~(\ref{RSch}) as formed in the equilibrium self-consistent potential well (see Fig.~\ref{pwell}).

\begin{figure}
\includegraphics[scale=1.1]{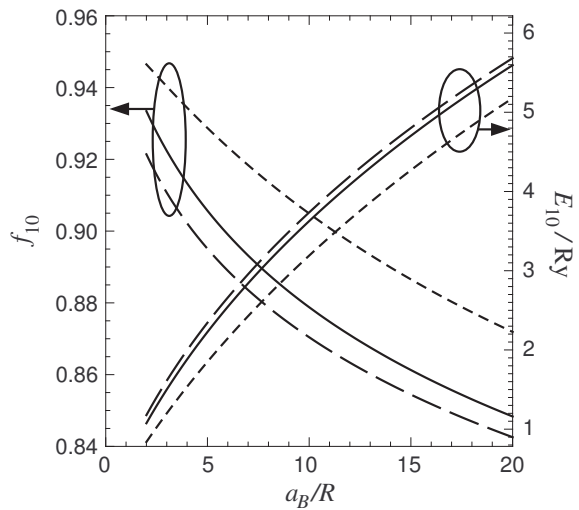}
\caption{\label{transosc} Oscillator strength and transition energy for the optical transition between $i=0$ and $i=1$ electronic eigenstates in the equlibrium potential well. Calculations done with different screening models are denoted as in Figs.~\ref{pszBE} and \ref{eratelev}.}
\end{figure}

For the optical absorption from the ground state, a Franck-Condon framework discussion requires electronic transition energies
\begin{equation}\label{trans1}
\Eio = \Ei - \Eo.
\end{equation}
The dominant dipole electronic transition turns out to be $i=0 \, \rightarrow \, i=1$, and Fig.~\ref{eratelev}(b) shows the relationship between the polaron binding energy $\bpol$ and the equilibrium value of $\Eoo$. (To compare, the ratio of these energies for 3D Pekar's polaron is about 0.8 \cite{appel}.) More detailed information on this transition for our 1D polaron is displayed in Fig.~\ref{transosc}; the most important conclusion from which is that the transition contains nearly 90\% of the total absorption strength across a wide range of the confinement parameter. This overall dominance is similar to the one taking place for 3D solvated electrons \cite{hydelec}.

Since Eq.~(\ref{RSch}) has a form of a standard continuum Schr\"{o}dinger equation, the overall absorption satisfies the optical sum rule familiar from atomic systems \cite{haugbook}:
$$
\sum_{i}\fio=1,
$$
where the oscillator strength of the transition to the $i$th state
\begin{equation}\label{trans2}
\fio = \frac{2\hbar^2}{m\Eio} \left|\pio \right|^2.
\end{equation}
In terms of the frequency $\om$-dependent oscillator strength density \cite{LL8},
the sum rule \textit{per one} polaron reads
\begin{equation}\label{sumRf}
\int_0^{\infty} \fom \, d\om \,=1,
\end{equation}
where, for sharp discrete transitions (\ref{trans1}),
\begin{equation}\label{fom}
\fom=\sum_{i}\fio\,\delta(\om-\omi), \ \ \ \omi = \Eio/\hbar.
\end{equation}
We will discuss in the next section how the thermal broadening of these transitions occurs resulting in a continuous $\fom$.

As Eq.~(\ref{trans2}) clearly indicates, the optical absorption in our case is due to the electric field polarized \textit{parallel} to the tube axis. The data in Fig.~\ref{transosc} has been derived by calculating matrix elements in Eq.~(\ref{trans2}) between the states in the equilibrium potential well (and the sum rule verified in all cases).

\section{Thermal broadening of the optical absorption}

In the Franck-Condon framework, the optical absorption from the polaron is realized by virtue of ``vertical'' electronic transitions between the adiabatic potential surfaces $\Etot_i$, meaning that the polarization state of the environment does not change while the transition takes place. This picture of \textit{slow} modulating dynamics  of the environment leads to what is called inhomogeneous broadening of the absorption lineshape (see Chapter 5 of Ref.~\cite{CET2004} for an extensive discussion) and is well justified for our application to the case of the orientational solvent polarization. Indeed, the parameter range of our interest and practical significance corresponds to the regime of $\bpol > \kB T > \hbar/\tal$, and, hence,
$$
\tal \gg \hbar\,(\kB T \bpol)^{-1/2},
$$
which is a condition of the applicability of the semi-classical Franck-Condon picture \cite{CET2004}. Here $\tal$ is the longitudinal relaxation time of the solvent defining the relevant time scale for the solvent orientational dynamics and equal to \cite{frobook,nitzan,fawcett}
$$ %\begin{equation}\label{deftauL}
\tau_L=(\epsi/\epss)\,\tad
$$ %\end{equation}
for the Debye solvent described by the $\om$-dependent dielectric function
\begin{equation}\label{deb1}
\epom = \epsi + \frac{\epss - \epsi}{1- i\om\tad}.
\end{equation}
Typical solvents are characterized by a wide range of $\tal$ ranging from fractions to tens of ps \cite{fawcett, YNGpol}.

Thermal fluctuations result in variations of the potential distribution $\potx$ in Eq.~(\ref{RSch}) from the equilibrium pattern $\pot (x)$ discussed in the previous section, thereby modulating electronic wave functions and transition energies (\ref{trans1}). Averaging over various fluctuations would thus lead to broadening of the discrete transitions in Eq.~(\ref{fom}). In the discussed semi-classical picture, the weighting of various fluctuations in the ground state is described by the Boltzmann factor associated with fluctuations of the \textit{total} adiabatic energy (\ref{ad1}) from the minimum equilibrium value (\ref{ad3}). Of course, fluctuations in the presence of the polaron are not the fluctuations of the solvent alone but rather combined fluctuations of the medium and localized electronic charge density.

One can get an insight into the nature of thermal fluctuations of the polaron shape by using explicit results derived in Ref.~\cite{GU_lowfreq} for \textit{small-amplitude} fluctuations in a simplified model of the charge carrier interacting with the Debye solvent (\ref{deb1}) with response function (\ref{Gsimp}). Within that model, shape fluctuations are governed by amplitudes $\ant$ in the expansion of the time-dependent electric potential $\potxt$ and self-localized carrier charge density $\denxt$ over the normal dielectric relaxation modes:
\begin{subequations}\label{ex12}
\begin{eqnarray}
\potxt & = & \pot (\xrelt) + \sumn \ant \potn (\xrelt), \ \ \ \ \label{ex1}\\
\denxt & = & \den (\xrelt) + \sumn \ant \denn (\xrelt), \ \ \ \
\label{ex2}
\end{eqnarray}
\end{subequations}
where the first terms in the r.~h.~s. of Eqs.~(\ref{ex12}) correspond to the equilibrium static patterns. Both equilibrium patterns and normal modes are centered around arbitrary polaron centroid positions $\ksit$. The dynamics of $\ksit$ results in the diffusion of the polaron as a whole and does not affect the optical absorption lineshape. The spatial patterns of the normal modes, $(\potn (x), \ \denn (x))$, follow from the solutions  of the generalized eigenvalue ($\lan$) problem:
\begin{subequations}\label{p4}
\begin{eqnarray}
\denn (x) & = & - \int d\xp L(x,\xp)\, \potn (\xp), \label{p4a} \\
\potn (x) & = & - \frac{\lan}{\est}\, \int d\xp G_0 (x-\xp)\,\denn (\xp),
\label{p4b}
\end{eqnarray}
\end{subequations}
where spatial kernel $L$ is determined by the response of the electronic subsystem (see Ref.~\cite{GU_lowfreq} for more detail). Equations (\ref{p4}) yield shape-modulating modes ($n \geq 1$) with growing eigenvalues $\lan > 1$ as well as the zero-frequency translational mode with $\lambda_0=1$, which is excluded in Eq.~(\ref{ex12}) in favor of the collective coordinate $\ksit$.

For our purposes here, it is convenient to choose the normalization of the modes as
\begin{equation}\label{p5}
-\int dx\, \denm (x)\, \potn (x) = \frac{1}{\lan}\,\delta_{nm},
\end{equation}
note that coefficients $\ant$ are then measured in units of the square root of energy. The equations of motion for so normalized modes read \cite{GU_lowfreq}
\begin{equation}\label{m4b}
\dot{a}_n +\an/\taln =   \fm_{n},
\end{equation}
where mode relaxation times $\taln$ are given by
\begin{equation}\label{m4bb}
\tal/\taln =1-1/\lan
\end{equation}
and perturbations $\fm_n$ by
\begin{equation}\label{m5b}
\fm_{n} (\xi,t)= - \lan \int dx \denn (x) F(x+\xi,t).
\end{equation}
As the eigenvalues $\lan$ grow with the mode index $n$, the relaxation times $\taln$ in (\ref{m4bb}) converge to the solvent longitudinal relaxation time $\tal$ (see \cite{GU_lowfreq} for an illustration of this convergence).

For thermal dielectric fluctuations, $F(x,t)$ in (\ref{m5b}) is a random process with a vanishing average and correlations established by ``bare" solvent fluctuations \cite{GU_lowfreq}:
\begin{equation}\label{c7}
\avg{F (x,t) F (\xp,\tp)} = \frac{2 \kB T}{\est\tal} \, G_0 (x-\xp)
\, \delta (t-\tp).
\end{equation}
It then follows from (\ref{p5}), (\ref{m5b}), and (\ref{c7}) that different modes $n$ and $m$ exhibit no dynamic correlations, and the only relevant (for the same centroid coordinate) non-zero correlation is
$$ %\begin{equation}\label{c8b}
\avg{\fm_n (\xi,t)\,\fm_n (\xi,\tp)} = 2 D \, \delta (t-\tp), \ \ \  D= \kB T / \tal,
$$ %\end{equation}
with the effective diffusion coefficient $D$ independent of the mode index with the chosen normalization.

The random dynamics described by Eq.~(\ref{m4b}) is thus the Ornstein--Uhlenbeck process \cite{stat2004}, which leads, at times $t \gg \taln$, to the Gaussian distribution of $\an$ with the variance established by
$$
\avg{\an^2} = D\taln:
$$
the probability
\begin{equation}\label{m4g}
P(\an) \propto \exp\left(-\frac{U_n (\an)}{\kB T}\right), \ \ \
U_n (\an) = \frac{\an^2}{2}\left(1 - \frac{1}{\lan} \right).
\end{equation}
Equation (\ref{m4g}) features the classic Boltzmann distribution for independent fluctuation modes with the effective potential energies $U_n (\an)$. The behavior of $U_n$ is consistent with the behavior (\ref{m4bb}) of relaxation times.

\begin{figure}
\includegraphics[scale=0.9]{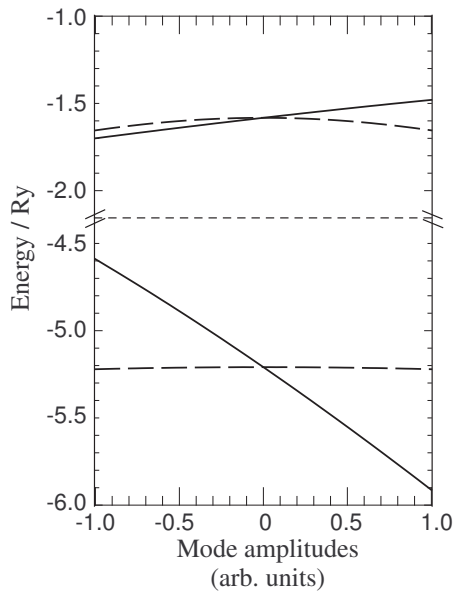}
\caption{\label{modecmp} Modulation of the two lowest electronic energy levels $\Eo$ and $E_1$ by the normal fluctuations modes, Eqs.~(\ref{ex12}), as a function of mode amplitudes in arbitrary units. Solid lines show the modulation by the even mode $n=1$ and long-dash lines by the odd mode $n=2$. The confinement parameter used for this illustration is $\aB/R=10$.}
\end{figure}

As the normal mode index $n$ grows, the interaction of the solvent polarization with the electronic density in the ground state generally decreases, likewise the modulation effect on the transition energy $\Eoo$ generally diminishes. The largest modulation effect occurs due to the even normal fluctuation mode $n=1$, corresponding to
the shallowest potential energy $U_1$. This modulation is illustrated in Fig.~\ref{modecmp} comparing the effects of two normal modes on the electronic energies $\Eo$ and $E_1$. It is evident that the major contribution comes from the modulation of the ground state level $\Eo$ by the $n=1$ mode. As discussed in Ref.~\cite{GU_lowfreq}, the spatial pattern of this mode corresponds qualitatively to the modulation of the polaron length $l$, we will accordingly call this the ``breathing'' mode.

While illuminating the character of the fluctuations and their modulation effects, the small-amplitude description turns out to be insufficient for realistic $T$ on the order of room temperature. In this paper we do not pursue a comprehensive analysis of the absorption lineshape that would then be required. Instead, we will provide a qualitative assessment of the effect of the polaron breathing using a simplistic single-parameter scaling description similar in the spirit to the approach used in a recent study \cite{ContScaling} of a continuum adiabatic model of 3D electron solvation. In this description, the fluctuations are restricted to modulate the electrostatic potential via a scaling ansatz of the equilibrium pattern:
\begin{equation}\label{rescaling}
\potx =c\,\pot \left(c\,x\right),
\end{equation}
where $c \geq 0$ is the scaling parameter. Its deviations from the value of $c=1$ results in variations of all quantities relevant for the optical absorption. It should be noted that the results of small deviations from $c=1$ turn out to be quite close to the results we derived with exact small-amplitude breathing fluctuations.

\begin{figure}
\includegraphics[scale=1.1]{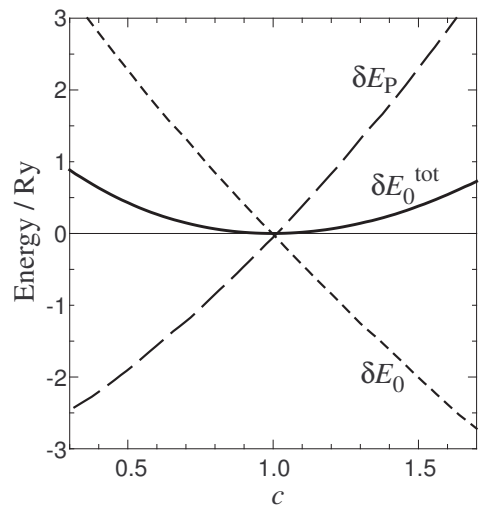}
\caption{\label{TEeq} Variations of the electronic energy $\delta\Eo$, polarization energy $\delta\Epol$ and the total adiabatic energy $\delta\Etot_0$ with the scaling parameter $c$. The confinement parameter used for this illustration is $\aB/R=10$.}
\end{figure}

\begin{figure}
\includegraphics[scale=1.1]{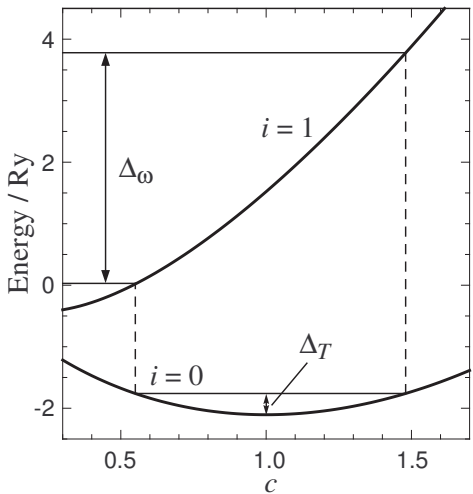}
\caption{\label{kTvTE} Total adiabatic energies $\Etot_0$ and $\Etot_1$ as functions of the scaling parameter $c$ calculated here for the confinement parameter $\aB/R=10$. The plot illustrates how fluctuations of small energy $\Delta_T$ translate into much larger variations $\Delta_{\omega}$ of the electronic transition energies.}
\end{figure}

Figure \ref{TEeq} displays variations of different energies $\delta E (c)= E(c) - E(c=1)$ with the scaling parameter $c$. It is evident that the shallow response of the total energy $\delta \Etot_0$ to fluctuations around the equilibrium is a result of a near compensation of much larger variations of the polarization energy $\delta \Epol$ and the purely electronic energy $\delta\Eo$, as indeed should be the case for the ground state adiabatic surface. It is this strong variability of $\Eo$ in response to the polaron breathing that has already been noted in Fig.~\ref{modecmp} and is largely responsible for the resulting enhanced broadening of the lineshape. The ``mechanism'' of the enhancement is also illustrated in Fig.~\ref{kTvTE} displaying two adiabatic energy surfaces, $\Etot_0$ and $\Etot_1$, between which the dominant optical transitions take place. The behavior of $\Etot_1$ here reflects large variations $\delta\Epol$ of the polarization energy without a comparable compensation of the electronic energy that occurs for $\Etot_0$. Figure \ref{kTvTE} shows how a population on $\Etot_0$ within a relatively small energetic broadening $\Delta_T$ translates into a much broader distribution $\Delta_{\omega}$ of the transition energies.

To model the absorption lineshape as a function of frequency $\omega$, we work with an ensemble average over various fluctuations represented by different values of the scaling parameter $c$ in Eq.~(\ref{rescaling}). The resulting oscillator strength density in Eq.~(\ref{sumRf}) is then a Boltzmann average of Eq.~(\ref{fom}) taking the form of
\begin{equation}\label{avintsty}
\fom= \int dc \, P(c) \, \sum_{i}\fio (c) \,\delta(\om-\omi (c)),
\end{equation}
where the probability
$$
P(c) = \exp\left(-\frac{\delta\Etot_0 (c)}{\kB T} \right) \left/ \int dc \, \exp\left(-\frac{\delta\Etot_0 (c)}{\kB T} \right) \right.
$$
determines the weighting of fluctuations on the ground state adiabatic surface $\Etot_0$. Both transition frequencies $\omi (c)$ and oscillator strengths $\fio (c)$ in Eq.~(\ref{avintsty}) are functions of the scaling parameter $c$.

\begin{figure}
\includegraphics[scale=1.1]{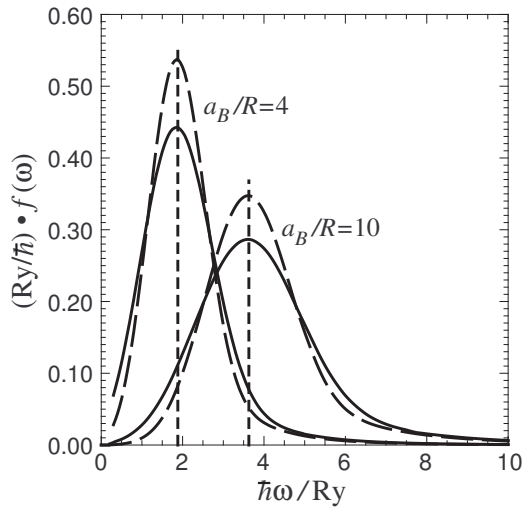}
\caption{\label{sclabs} The model absorption lineshapes calculated as described in the text for $\aB/R=4$ and $\aB/R=10$ and for two different temperatures: $\kB T =0.2\,\Ry$ (long-dash curves) and $\kB T =0.3\,\Ry$ (solid curves). Vertical short-dash lines show the positions of the respective electronic transition energies $\Eoo$ in the equilibrium configurations. The low-frequency tail of $\fom$ for the case of $\aB/R=4$ is not shown completely as the computational scheme used becomes inadequate.}
\end{figure}

Figure \ref{sclabs} shows examples of the absorption lineshapes calculated with Eq.~(\ref{avintsty}) for two values of the confinement parameter and two different values of temperature $T$. The results are broad peaks around the respective equilibrium transitions frequencies $\Eoo$. The broadening of the absorption lines, as we discussed, is strongly enhanced, roughly an order of magnitude larger than the thermal energies in our examples. In this respect, it is similar to the results known for 3D solvated electrons \cite{hydelec}. It is this large and confinement-parameter-dependent broadening that is the emphasis of Fig.~\ref{sclabs} rather than details of the model lineshapes.

\section{Conclusions}

As a result of solvation by the surrounding sluggish polar medium, excess charge carriers on 1D semiconductor nanostructures can undergo self-localization into 1D polarons whose properties are quite different from band carriers. In an effort to establish experimentally testable signatures of the polaron formation, in this paper we examined the local optical absorption from solvation-induced 1D polarons within a simplified single-particle adiabatic framework for tubular structures.

The calculations indicate that a new broad absorption feature is expected, the peak energy of which is comparable with (but larger than) the binding energy of the polaron and whose width is much larger than the thermal energy. This feature arises mostly due to the electronic transition between two lowest-energy localized electronic levels formed in the polarization potential well. We identified the polaron length fluctuations (``breathing'' mode) as the major source of the broadening.

Similarly to the enhanced binding of 1D excitons \cite{haugbook}, the binding of 1D polarons is stronger than in higher dimensions, and we explored polaron structure and optical transitions as a function of the confinement parameter $\aB/R$, where $\aB$ is the 3D Bohr radius (\ref{defaB}) and $R$ the radius of the tubular structure. Of particular relevance for the optical absorption is the equilibrium transition energy $\Eoo$, which approximately determines the position of the absorption peak (Fig.~\ref{sclabs}). Within the range shown in Fig.~\ref{transosc}, the scaling of this energy may be \textit{approximated} as
\begin{equation}\label{sc1}
\Eoo / \Ry \propto (\aB/R)^{0.7}.
\end{equation}
Note that the relation between $\Eoo$ and the binding energy $\bpol$ also changes with the confinement parameter (Fig.~\ref{eratelev}). Together with the parametric dependence (\ref{defaB}) of the Rydberg energy $\Ry$, Eq.~(\ref{sc1}) allows to assess how the position of the absorption peak scales with system parameters (compare to scaling of excitons in SWCNTs \cite{pedersen,tersscaling}). It follows then that the effective dielectric constant $\est$ of the solvent affects it as
$$
\Eoo \propto (\est)^{-1.3}.
$$
If the effective mass $m$ of the carrier was independent of the tube radius $R$, then (\ref{sc1}) would result in
$$
\Eoo \propto m^{0.3}.
$$
In SWCNTs, however, $m$ does depend on tube radius, and, for not very small radius tubes, this dependence is roughly $m \propto 1/R$ \cite{pedersen,NTEffectMass}. In this regime, the confinement parameter becomes practically independent of $R$, and the transition energy scales as
$$
\Eoo \propto m \propto R^{-1}.
$$
Numerical estimates provided in the Introduction suggest that the absorption peak may be expected at energies on the order of 0.1 eV, subject, of course, to variations of the effective system parameters.

To reiterate, illustrative results in this paper have been derived using simplified model considerations. More accurate treatments, particularly of temperature effects, may be needed for detailed comparisons with experimental data when it becomes available.

We gratefully acknowledge support from the Collaborative U.~T.~Dallas -- SPRING Research and Nanotechnology Transfer Program.

\bibliography{fluctuations}

\begin{thebibliography}{35}
\expandafter\ifx\csname natexlab\endcsname\relax\def\natexlab#1{#1}\fi
\expandafter\ifx\csname bibnamefont\endcsname\relax
  \def\bibnamefont#1{#1}\fi
\expandafter\ifx\csname bibfnamefont\endcsname\relax
  \def\bibfnamefont#1{#1}\fi
\expandafter\ifx\csname citenamefont\endcsname\relax
  \def\citenamefont#1{#1}\fi
\expandafter\ifx\csname url\endcsname\relax
  \def\url#1{\texttt{#1}}\fi
\expandafter\ifx\csname urlprefix\endcsname\relax\def\urlprefix{URL }\fi
\providecommand{\bibinfo}[2]{#2}
\providecommand{\eprint}[2][]{\url{#2}}

\bibitem[{\citenamefont{Licht}(2002)}]{SCelectrodes}
\bibinfo{editor}{\bibfnamefont{S.}~\bibnamefont{Licht}}, ed.,
  \emph{\bibinfo{title}{Semiconductor Electrodes and Photoelectrochemistry}}
  (\bibinfo{publisher}{Wiley-VCH}, \bibinfo{address}{Weinheim, Germany},
  \bibinfo{year}{2002}).

\bibitem[{\citenamefont{Kamat}(2007)}]{kamat07}
\bibinfo{author}{\bibfnamefont{P.~V.} \bibnamefont{Kamat}},
  \bibinfo{journal}{J. Phys. Chem. C} \textbf{\bibinfo{volume}{111}},
  \bibinfo{pages}{2834} (\bibinfo{year}{2007}).

\bibitem[{\citenamefont{Law et~al.}(2005)\citenamefont{Law, Greene, Johnson,
  Saykally, and Yang}}]{pyang05}
\bibinfo{author}{\bibfnamefont{M.}~\bibnamefont{Law}},
  \bibinfo{author}{\bibfnamefont{L.~E.} \bibnamefont{Greene}},
  \bibinfo{author}{\bibfnamefont{J.~C.} \bibnamefont{Johnson}},
  \bibinfo{author}{\bibfnamefont{R.}~\bibnamefont{Saykally}}, \bibnamefont{and}
  \bibinfo{author}{\bibfnamefont{P.}~\bibnamefont{Yang}},
  \bibinfo{journal}{Nature Materials} \textbf{\bibinfo{volume}{4}},
  \bibinfo{pages}{455} (\bibinfo{year}{2005}).

\bibitem[{\citenamefont{Basko and Conwell}(2002)}]{basko}
\bibinfo{author}{\bibfnamefont{D.~M.} \bibnamefont{Basko}} \bibnamefont{and}
  \bibinfo{author}{\bibfnamefont{E.~M.} \bibnamefont{Conwell}},
  \bibinfo{journal}{Phys. Rev. Lett.} \textbf{\bibinfo{volume}{88}},
  \bibinfo{pages}{098102} (\bibinfo{year}{2002}).

\bibitem[{\citenamefont{Gartstein}(2006)}]{YNGpol}
\bibinfo{author}{\bibfnamefont{Y.~N.} \bibnamefont{Gartstein}},
  \bibinfo{journal}{Phys. Lett. A} \textbf{\bibinfo{volume}{349}},
  \bibinfo{pages}{377} (\bibinfo{year}{2006}).

\bibitem[{\citenamefont{Gartstein et~al.}(2007)\citenamefont{Gartstein,
  Bustamante, and {Ortega Castillo}}}]{polcylinder}
\bibinfo{author}{\bibfnamefont{Y.~N.} \bibnamefont{Gartstein}},
  \bibinfo{author}{\bibfnamefont{T.~D.} \bibnamefont{Bustamante}},
  \bibnamefont{and} \bibinfo{author}{\bibfnamefont{S.}~\bibnamefont{{Ortega
  Castillo}}}, \bibinfo{journal}{J. Phys.: Condens. Matter}
  \textbf{\bibinfo{volume}{19}}, \bibinfo{pages}{156210}
  (\bibinfo{year}{2007}).

\bibitem[{\citenamefont{Mayo and Gartstein}(2008)}]{MG_abinitio1}
\bibinfo{author}{\bibfnamefont{M.~L.} \bibnamefont{Mayo}} \bibnamefont{and}
  \bibinfo{author}{\bibfnamefont{Y.~N.} \bibnamefont{Gartstein}},
  \bibinfo{journal}{Phys. Rev. B} \textbf{\bibinfo{volume}{78}},
  \bibinfo{pages}{073402} (\bibinfo{year}{2008}).

\bibitem[{\citenamefont{Gartstein and Usery}(2008)}]{GU_lowfreq}
\bibinfo{author}{\bibfnamefont{Y.~N.} \bibnamefont{Gartstein}}
  \bibnamefont{and} \bibinfo{author}{\bibfnamefont{G.~L.} \bibnamefont{Usery}},
  \bibinfo{journal}{Phys. Lett. A}  (\bibinfo{year}{2008}),
  \bibinfo{note}{doi:10.1016/j.physleta.2008.07.031}.

\bibitem[{\citenamefont{Kuper and Whitfield}(1963)}]{polarons1}
\bibinfo{editor}{\bibfnamefont{C.~G.} \bibnamefont{Kuper}} \bibnamefont{and}
  \bibinfo{editor}{\bibfnamefont{G.~D.} \bibnamefont{Whitfield}}, eds.,
  \emph{\bibinfo{title}{Polarons and excitons}} (\bibinfo{publisher}{Plenum},
  \bibinfo{address}{New York}, \bibinfo{year}{1963}).

\bibitem[{\citenamefont{Devreese}(1972)}]{polarons2}
\bibinfo{editor}{\bibfnamefont{J.~T.} \bibnamefont{Devreese}}, ed.,
  \emph{\bibinfo{title}{Polarons in ionic crystals and polar semiconductors}}
  (\bibinfo{publisher}{North Holland}, \bibinfo{address}{Amsterdam},
  \bibinfo{year}{1972}).

\bibitem[{\citenamefont{Kuznetsov}(1995)}]{CTbook}
\bibinfo{author}{\bibfnamefont{A.~M.} \bibnamefont{Kuznetsov}},
  \emph{\bibinfo{title}{Charge Transfer in Physics, Chemistry and Biology}}
  (\bibinfo{publisher}{Gordon and Breach}, \bibinfo{address}{Luxembourg},
  \bibinfo{year}{1995}).

\bibitem[{\citenamefont{Ferradini and {Jay-Gerin}}(1991)}]{ferra91}
\bibinfo{editor}{\bibfnamefont{C.}~\bibnamefont{Ferradini}} \bibnamefont{and}
  \bibinfo{editor}{\bibfnamefont{J.~P.} \bibnamefont{{Jay-Gerin}}}, eds.,
  \emph{\bibinfo{title}{Excess Electrons in Dielectric Media}}
  (\bibinfo{publisher}{CRC Press}, \bibinfo{address}{Boca Raton},
  \bibinfo{year}{1991}).

\bibitem[{\citenamefont{Nitzan}(2006)}]{nitzan}
\bibinfo{author}{\bibfnamefont{A.}~\bibnamefont{Nitzan}},
  \emph{\bibinfo{title}{Chemical Dynamics in Condensed Phases}}
  (\bibinfo{publisher}{Oxford}, \bibinfo{address}{New York},
  \bibinfo{year}{2006}).

\bibitem[{\citenamefont{Avouris and Chen}(2006)}]{NTelectronics}
\bibinfo{author}{\bibfnamefont{P.}~\bibnamefont{Avouris}} \bibnamefont{and}
  \bibinfo{author}{\bibfnamefont{J.}~\bibnamefont{Chen}},
  \bibinfo{journal}{Materials Today} \textbf{\bibinfo{volume}{9}},
  \bibinfo{pages}{46} (\bibinfo{year}{2006}).

\bibitem[{\citenamefont{Pennington and Goldsman}(2005)}]{NTEffectMass}
\bibinfo{author}{\bibfnamefont{G.}~\bibnamefont{Pennington}} \bibnamefont{and}
  \bibinfo{author}{\bibfnamefont{N.}~\bibnamefont{Goldsman}},
  \bibinfo{journal}{Phys. Rev. B} \textbf{\bibinfo{volume}{71}},
  \bibinfo{pages}{205318} (\bibinfo{year}{2005}).

\bibitem[{\citenamefont{Li et~al.}(2006)\citenamefont{Li, Qian, Xiang, and
  Lieber}}]{NWelectronics}
\bibinfo{author}{\bibfnamefont{Y.}~\bibnamefont{Li}},
  \bibinfo{author}{\bibfnamefont{F.}~\bibnamefont{Qian}},
  \bibinfo{author}{\bibfnamefont{J.}~\bibnamefont{Xiang}}, \bibnamefont{and}
  \bibinfo{author}{\bibfnamefont{C.~M.} \bibnamefont{Lieber}},
  \bibinfo{journal}{Materials Today} \textbf{\bibinfo{volume}{9}},
  \bibinfo{pages}{18} (\bibinfo{year}{2006}).

\bibitem[{\citenamefont{Conwell and Basko}(2006)}]{CBDrag}
\bibinfo{author}{\bibfnamefont{E.~M.} \bibnamefont{Conwell}} \bibnamefont{and}
  \bibinfo{author}{\bibfnamefont{D.~M.} \bibnamefont{Basko}},
  \bibinfo{journal}{J. Phys. Chem. B} \textbf{\bibinfo{volume}{110}},
  \bibinfo{pages}{23603} (\bibinfo{year}{2006}).

\bibitem[{\citenamefont{O'Connell et~al.}(2005)\citenamefont{O'Connell,
  Eibergen, and Doorn}}]{chirsel}
\bibinfo{author}{\bibfnamefont{M.~J.} \bibnamefont{O'Connell}},
  \bibinfo{author}{\bibfnamefont{E.~E.} \bibnamefont{Eibergen}},
  \bibnamefont{and} \bibinfo{author}{\bibfnamefont{S.~K.} \bibnamefont{Doorn}},
  \bibinfo{journal}{Nature Materials} \textbf{\bibinfo{volume}{4}},
  \bibinfo{pages}{412} (\bibinfo{year}{2005}).

\bibitem[{\citenamefont{Choi and Strano}(2007)}]{CNTsolvchrom}
\bibinfo{author}{\bibfnamefont{J.~H.} \bibnamefont{Choi}} \bibnamefont{and}
  \bibinfo{author}{\bibfnamefont{M.~S.} \bibnamefont{Strano}},
  \bibinfo{journal}{Appl. Phys. Lett.} \textbf{\bibinfo{volume}{90}},
  \bibinfo{pages}{223114} (\bibinfo{year}{2007}).

\bibitem[{\citenamefont{Haug and Koch}(2004)}]{haugbook}
\bibinfo{author}{\bibfnamefont{H.}~\bibnamefont{Haug}} \bibnamefont{and}
  \bibinfo{author}{\bibfnamefont{S.~W.} \bibnamefont{Koch}},
  \emph{\bibinfo{title}{Quantum theory of the optical and electronic properties
  of semiconductors}} (\bibinfo{publisher}{World Scientific},
  \bibinfo{address}{New Jersey}, \bibinfo{year}{2004}).

\bibitem[{\citenamefont{Pekar}(1946)}]{pekar1}
\bibinfo{author}{\bibfnamefont{S.~I.} \bibnamefont{Pekar}},
  \bibinfo{journal}{Zh. Eksp. Teor. Fiz.} \textbf{\bibinfo{volume}{16}},
  \bibinfo{pages}{335, 341} (\bibinfo{year}{1946}).

\bibitem[{\citenamefont{Appel}(1968)}]{appel}
\bibinfo{author}{\bibfnamefont{J.}~\bibnamefont{Appel}}, in
  \emph{\bibinfo{booktitle}{Solid State Physics}}, edited by
  \bibinfo{editor}{\bibfnamefont{F.}~\bibnamefont{Seitz}},
  \bibinfo{editor}{\bibfnamefont{D.}~\bibnamefont{Turnbull}}, \bibnamefont{and}
  \bibinfo{editor}{\bibfnamefont{H.}~\bibnamefont{Ehrenreich}}
  (\bibinfo{publisher}{Academic}, \bibinfo{address}{New York},
  \bibinfo{year}{1968}), vol.~\bibinfo{volume}{21}, p. \bibinfo{pages}{193}.

\bibitem[{\citenamefont{Wu et~al.}(1985)\citenamefont{Wu, Peeters, and
  Devreese}}]{pol2d}
\bibinfo{author}{\bibfnamefont{X.}~\bibnamefont{Wu}},
  \bibinfo{author}{\bibfnamefont{F.~M.} \bibnamefont{Peeters}},
  \bibnamefont{and} \bibinfo{author}{\bibfnamefont{J.~T.}
  \bibnamefont{Devreese}}, \bibinfo{journal}{Phys. Rev. B}
  \textbf{\bibinfo{volume}{31}}, \bibinfo{pages}{3420} (\bibinfo{year}{1985}).

\bibitem[{\citenamefont{Fawcett}(2004)}]{fawcett}
\bibinfo{author}{\bibfnamefont{W.~R.} \bibnamefont{Fawcett}},
  \emph{\bibinfo{title}{Liquids, solutions and interfaces}}
  (\bibinfo{publisher}{Oxford}, \bibinfo{address}{Oxford},
  \bibinfo{year}{2004}).

\bibitem[{\citenamefont{Schnitker et~al.}(1988)\citenamefont{Schnitker,
  Motakabbir, Rossky, and Friesner}}]{hydelec}
\bibinfo{author}{\bibfnamefont{J.}~\bibnamefont{Schnitker}},
  \bibinfo{author}{\bibfnamefont{K.}~\bibnamefont{Motakabbir}},
  \bibinfo{author}{\bibfnamefont{P.~J.} \bibnamefont{Rossky}},
  \bibnamefont{and} \bibinfo{author}{\bibfnamefont{R.~A.}
  \bibnamefont{Friesner}}, \bibinfo{journal}{Phys. Rev. Lett.}
  \textbf{\bibinfo{volume}{60}}, \bibinfo{pages}{456} (\bibinfo{year}{1988}).

\bibitem[{\citenamefont{May and K\"{u}hn}(2004)}]{CET2004}
\bibinfo{author}{\bibfnamefont{V.}~\bibnamefont{May}} \bibnamefont{and}
  \bibinfo{author}{\bibfnamefont{O.}~\bibnamefont{K\"{u}hn}},
  \emph{\bibinfo{title}{Charge and Energy Transfer Dynamics in Molecular
  Systems}} (\bibinfo{publisher}{Wiley-VCH}, \bibinfo{address}{Weinheim,
  Germany}, \bibinfo{year}{2004}).

\bibitem[{\citenamefont{Ando}(1997)}]{ando}
\bibinfo{author}{\bibfnamefont{T.}~\bibnamefont{Ando}}, \bibinfo{journal}{J.
  Phys. Soc. Japan} \textbf{\bibinfo{volume}{66}}, \bibinfo{pages}{1066}
  (\bibinfo{year}{1997}).

\bibitem[{\citenamefont{Jackson}(1998)}]{jackson}
\bibinfo{author}{\bibfnamefont{J.~D.} \bibnamefont{Jackson}},
  \emph{\bibinfo{title}{Classical electrodynamics}}
  (\bibinfo{publisher}{Wiley}, \bibinfo{address}{New York},
  \bibinfo{year}{1998}).

\bibitem[{\citenamefont{Jiang et~al.}(2007)\citenamefont{Jiang, Saito,
  Samsonidze, Jorio, Chou, Dresslhaus, and Dresslhaus}}]{jiang2007}
\bibinfo{author}{\bibfnamefont{J.}~\bibnamefont{Jiang}},
  \bibinfo{author}{\bibfnamefont{R.}~\bibnamefont{Saito}},
  \bibinfo{author}{\bibfnamefont{G.~G.} \bibnamefont{Samsonidze}},
  \bibinfo{author}{\bibfnamefont{A.}~\bibnamefont{Jorio}},
  \bibinfo{author}{\bibfnamefont{S.~G.} \bibnamefont{Chou}},
  \bibinfo{author}{\bibfnamefont{G.}~\bibnamefont{Dresslhaus}},
  \bibnamefont{and} \bibinfo{author}{\bibfnamefont{M.~S.}
  \bibnamefont{Dresslhaus}}, \bibinfo{journal}{Phys. Rev. B}
  \textbf{\bibinfo{volume}{75}}, \bibinfo{pages}{035407}
  (\bibinfo{year}{2007}).

\bibitem[{\citenamefont{Landau and Lifshitz}(1984)}]{LL8}
\bibinfo{author}{\bibfnamefont{L.~D.} \bibnamefont{Landau}} \bibnamefont{and}
  \bibinfo{author}{\bibfnamefont{E.~M.} \bibnamefont{Lifshitz}},
  \emph{\bibinfo{title}{Electrodynamics of Continuous Media}}
  (\bibinfo{publisher}{Butterworth-Heinemann}, \bibinfo{address}{Oxford},
  \bibinfo{year}{1984}).

\bibitem[{\citenamefont{Fr\"{o}hlich}(1949)}]{frobook}
\bibinfo{author}{\bibfnamefont{H.}~\bibnamefont{Fr\"{o}hlich}},
  \emph{\bibinfo{title}{Theory of dielectrics}}
  (\bibinfo{publisher}{Clarendon}, \bibinfo{address}{Oxford},
  \bibinfo{year}{1949}).

\bibitem[{\citenamefont{{Le Bellac} et~al.}(2004)\citenamefont{{Le Bellac},
  Mortessagne, and Batrouni}}]{stat2004}
\bibinfo{author}{\bibfnamefont{M.}~\bibnamefont{{Le Bellac}}},
  \bibinfo{author}{\bibfnamefont{F.}~\bibnamefont{Mortessagne}},
  \bibnamefont{and} \bibinfo{author}{\bibfnamefont{G.~G.}
  \bibnamefont{Batrouni}}, \emph{\bibinfo{title}{Equilibrium and
  Non-Equilibrium Statistical Thermodynamics}} (\bibinfo{publisher}{Cambridge
  University Press}, \bibinfo{address}{Cambridge}, \bibinfo{year}{2004}).

\bibitem[{\citenamefont{Zharikov and Fischer}(2006)}]{ContScaling}
\bibinfo{author}{\bibfnamefont{A.~A.} \bibnamefont{Zharikov}} \bibnamefont{and}
  \bibinfo{author}{\bibfnamefont{S.~F.} \bibnamefont{Fischer}},
  \bibinfo{journal}{J. Chem. Phys.} \textbf{\bibinfo{volume}{124}},
  \bibinfo{pages}{054506} (\bibinfo{year}{2006}).

\bibitem[{\citenamefont{Pedersen}(2003)}]{pedersen}
\bibinfo{author}{\bibfnamefont{T.~G.} \bibnamefont{Pedersen}},
  \bibinfo{journal}{Phys. Rev. B} \textbf{\bibinfo{volume}{67}},
  \bibinfo{pages}{073401} (\bibinfo{year}{2003}).

\bibitem[{\citenamefont{Perebeinos et~al.}(2004)\citenamefont{Perebeinos,
  Tersoff, and Avouris}}]{tersscaling}
\bibinfo{author}{\bibfnamefont{V.}~\bibnamefont{Perebeinos}},
  \bibinfo{author}{\bibfnamefont{J.}~\bibnamefont{Tersoff}}, \bibnamefont{and}
  \bibinfo{author}{\bibfnamefont{P.}~\bibnamefont{Avouris}},
  \bibinfo{journal}{Phys. Rev. Lett.} \textbf{\bibinfo{volume}{92}},
  \bibinfo{pages}{257402} (\bibinfo{year}{2004}).

\end{thebibliography}

\end{document}